\documentclass[conference]{IEEEtran}
\IEEEoverridecommandlockouts

\usepackage{cite}
\usepackage[english]{babel}
\usepackage[T1]{fontenc}
\usepackage{lmodern}
\usepackage{mathrsfs}
\usepackage{mathptmx}  % Times with proper math support (replaces times package)
\usepackage{array}
\usepackage{float}
\usepackage{textcomp}
\usepackage{multirow}
\usepackage{stackrel}
\usepackage{graphicx}  % Remove duplicate
\usepackage{booktabs}
\usepackage{setspace}
\usepackage[export]{adjustbox}
\usepackage{algorithm}
\usepackage{caption}
\usepackage{subcaption}
\usepackage{algpseudocode}
\usepackage{enumerate}
\usepackage{xcolor}  % xcolor supersedes color package
\usepackage{amssymb}
\usepackage{amsmath,amssymb,amsfonts}
\usepackage{amsmath,amssymb,amsfonts}
\usepackage[mathcal]{eucal}
\def\BibTeX{{\rm B\kern-.05em{\sc i\kern-.025em b}\kern-.08em
    T\kern-.1667em\lower.7ex\hbox{E}\kern-.125emX}}

\begin{document}
\title{EcoFL: Resource Allocation for Energy-Efficient Federated Learning in Multi-RAT ORAN Networks}
\author{
  Abdelaziz Salama\textsuperscript{$\star$} \quad Mohammed M. H. Qazzaz\textsuperscript{$\star$} \quad Syed Danial Ali Shah\textsuperscript{$\star$} \quad Maryam Hafeez\textsuperscript{$\star$}
  \\ \quad Syed Ali Zaidi\textsuperscript{$\star$} \quad Hamed Ahmadi\textsuperscript{$\star\star$}\\ 
  \textsuperscript{$\star$}School of Electrical and Electronic Engineering, University of Leeds, Leeds, UK\\
  \textsuperscript{$\star\star$}School of Electrical and Electronic Engineering, University of York, Leeds, UK\\
  \textsuperscript{$\star$}Corresponding author: Abdelaziz Salama (A.M.Salama@Leeds.ac.uk)
}

\maketitle

\begin{abstract}
Federated Learning (FL) enables distributed model training on edge devices while preserving data privacy. However, FL deployments in wireless networks face significant challenges, including communication overhead, unreliable connectivity, and high energy consumption, particularly in dynamic environments. This paper proposes EcoFL, an integrated FL framework that leverages the Open Radio Access Network (ORAN) architecture with multiple Radio Access Technologies (RATs) to enhance communication efficiency and ensure robust FL operations. EcoFL implements a two-stage optimisation approach: an RL-based rApp for dynamic RAT selection that balances energy efficiency with network performance, and a CNN-based xApp for near real-time resource allocation with adaptive policies. This coordinated approach significantly enhances communication resilience under fluctuating network conditions. Experimental results demonstrate competitive FL model performance with 19\% lower power consumption compared to baseline approaches, highlighting substantial potential for scalable, energy-efficient collaborative learning applications.
\end{abstract}


\section{Introduction}

The rapid proliferation of edge devices, including smartphones and IoT sensors, has driven the growth of data-intensive applications. Federated Learning (FL) has emerged as a crucial distributed machine learning paradigm, enabling collaborative model training directly on edge devices while preserving data privacy \cite{mcmahan2017communicationefficient}. However, FL deployments in wireless networks face critical challenges, including substantial communication overhead, unreliable connectivity, and excessive energy consumption, particularly in dynamic and heterogeneous environments \cite{gecer2024federated}.

While existing studies have explored FL mechanisms for various network applications and Quality of Service (QoS) requirements \cite{bouzinis2021wireless,li2023survey}, optimising FL performance under communication constraints and network limitations remains challenging, especially in dynamic environments where efficient resource management is crucial for FL users.

To address these limitations, we propose \textbf{EcoFL}, a novel FL framework that dynamically optimises communication efficiency and energy consumption by leveraging Open Radio Access Network (ORAN) architecture \cite{qiao2024ai,polese2023understanding}. ORAN's modular and open architecture enables intelligent control over Radio Access Technologies (RATs), facilitating adaptive resource allocation. Unlike conventional FL implementations constrained by single, often congested RATs, EcoFL intelligently selects the most efficient RAT for each FL client based on real-time network conditions, reducing communication bottlenecks while enhancing reliability and minimising power consumption.

\begin{figure}[t]
    \centering
    \includegraphics[width=1\linewidth, height=0.29\textheight]{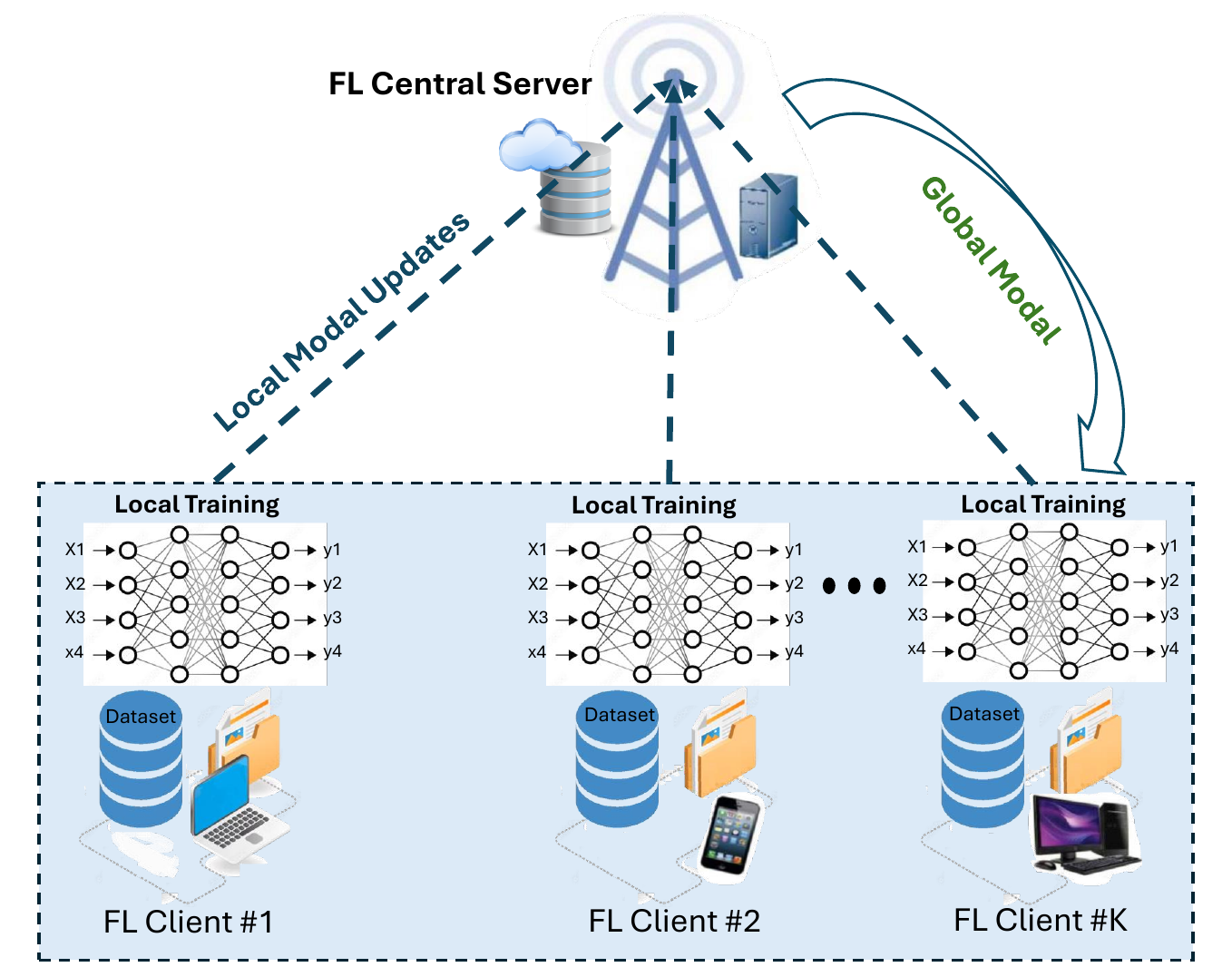}
    \caption{Conventional Federated Learning architecture showing traditional limitations.}
    \label{fig:tradional FL}
\end{figure}

The primary contributions of this paper are:
\begin{itemize}
    \item \textbf{Dynamic RAT Selection}: An adaptive RL-based mechanism for optimal RAT selection that ensures efficient communication while minimising latency and energy consumption.
    \item \textbf{Energy-Aware Resource Management}: CNN-based xApps within ORAN that intelligently allocate network resources, reducing power consumption without compromising FL performance.
    \item \textbf{Enhanced FL Efficiency}: A comprehensive framework that improves FL processes by minimising communication overhead, boosting scalability, and ensuring robust resource allocation.
\end{itemize}

EcoFL integrates FL with ORAN's adaptive capabilities, enabling intelligent load balancing across different RATs. The ORAN-based rApp/xApp framework continuously monitors network conditions, analyses traffic patterns, and optimises resource allocation to maintain high FL performance with reduced energy costs.

The remainder of this paper is structured as follows: Section \ref{Sec: Related work} reviews related works and research challenges. Section \ref{Sec: System Model} presents the system model and proposed optimisation strategy. Section \ref{sec: Simulation and results} evaluates EcoFL's performance against baseline approaches. Finally, Section \ref{sec: conclusion} concludes the paper.

\section{Related Works}
\label{Sec: Related work}

Recent research has extensively explored FL integration with ORAN architectures. McMahan et al. \cite{mcmahan2017communication} introduced the foundational concepts of FL for distributed deep learning by selecting the client subset to enhance privacy. However, this approach presents challenges in model performance and stability, particularly in complex applications where global aggregation may experience slower convergence due to limited client participation.

Various FL variants have been developed for mobile edge-cloud architectures \cite{chen2020joint,yang2020energy,dinh2020federated, mo2021energy}. Chen et al. \cite{chen2020joint} and Mo et al. \cite{mo2021energy} optimise energy efficiency through joint optimisation under computational constraints. In contrast, Yang et al. \cite{yang2020energy} incorporate latency constraints to minimise device energy consumption. Dinh et al. \cite{dinh2020federated} investigate FL in wireless environments using synchronous communication. Recent efforts include compression models to mitigate communication overhead \cite{pei2024dual}, although their impact on convergence requires further clarification.

Emerging applications, such as anomaly detection in 6G-enabled IoV using FL within ORAN networks \cite{driss2024federated, polese2023understanding}, demonstrate growing interest in practical FL deployments. However, a significant research gap persists: efficient network resource allocation within dynamic FL networks, particularly for prioritising FL training without compromising network stability or critical services.

Most existing works assume static network conditions or inadequately address how communication challenges and network dynamics lead to FL client outages due to unmanaged resource utilisation. Our work bridges this gap by proposing a novel ORAN-integrated approach that dynamically optimises RAT selection and power consumption specifically for FL networks, ensuring reliable training and stable learning outcomes.

\section{System Model}
\label{Sec: System Model}

This section presents a system model integrating ORAN architecture with FL, leveraging ORAN's modular flexibility for efficient multi-RAT communication. We investigate FL deployment in cellular networks through ORAN, where $B$ base stations serve $K$ users simultaneously. The core network processes aggregated data and models, functioning as the central server in our FL framework, responsible for global model updates and training coordination across distributed base stations while maintaining data privacy \cite{salama2023decentralized}.

\begin{figure}[t]
    \centering
    \includegraphics[width=1\linewidth, height=0.35\textheight]{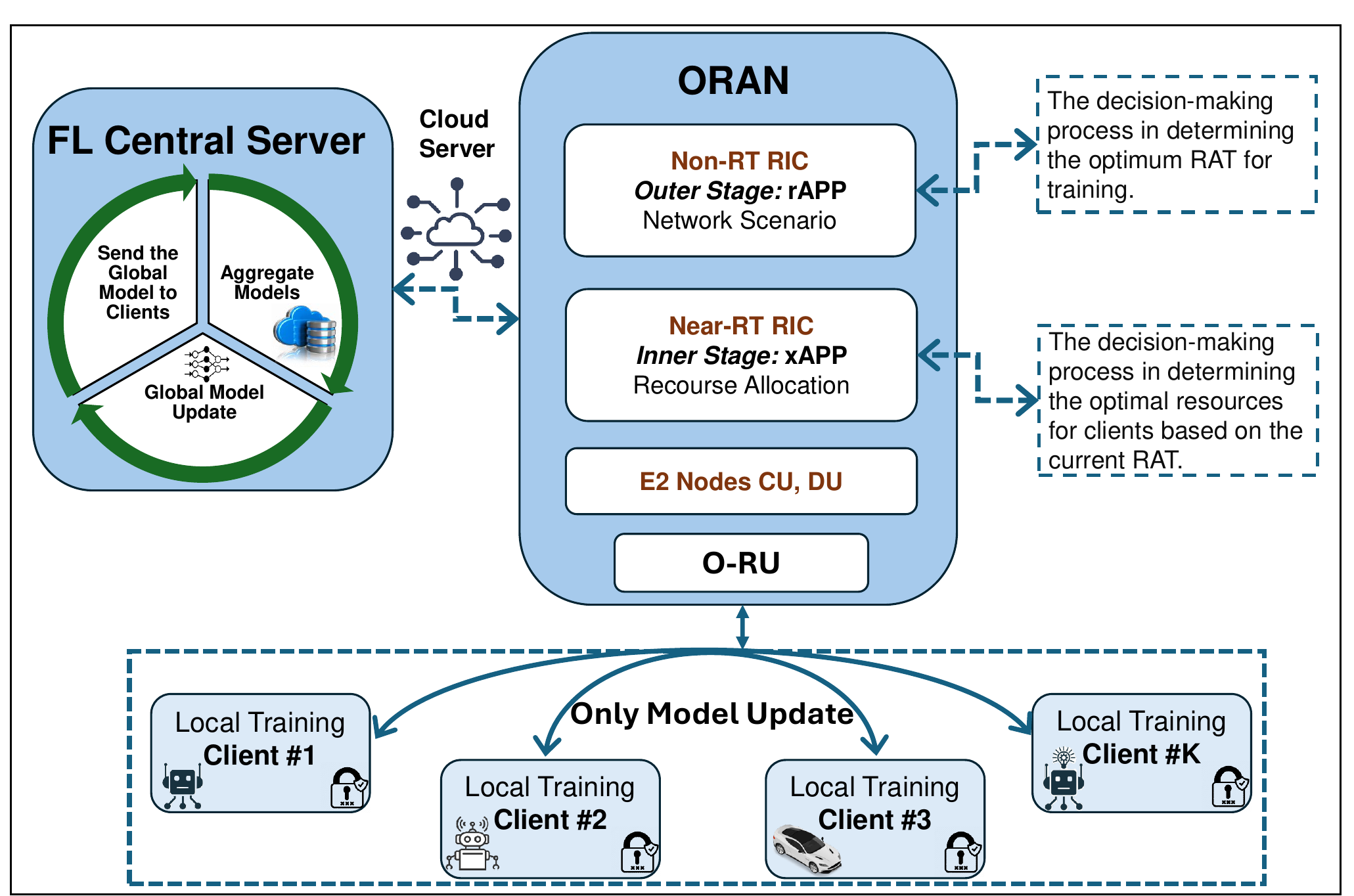}
    \caption{The proposed EcoFL network architecture.}
    \label{fig:ORAN FL}
\end{figure}

Traffic is categorised into different classes with users connected through multiple RATs via dual connectivity. There are $R$ RAT types $(r_1, r_2, \ldots, r_R)$, where each $r$ represents a specific technology (e.g., LTE, 5G NR). Figure \ref{fig:ORAN FL} illustrates the wireless system layout, showing RAN Intelligent Controllers (RICs): Non-Real-Time (Non-RT) RIC and Near-Real-Time (Near-RT) RIC, designed to host rApps and xApps for control and optimisation across different time scales.

\subsection{ORAN and Multi-RAT Network Configuration}

Each user device represents a distributed node within an ORAN-based network configured with RIC to facilitate data transmission over various RATs. ORAN's modular structure disaggregates traditional gNB components into three units: Open Radio Unit (O-RU), Open Distributed Unit (O-DU), and Open Centralised Unit (O-CU) \cite{polese2023understanding}, strategically deployed across base stations and edge cloud for efficient FL parameter transmission.

Using ORAN capabilities \cite{habib2024transformer}, our system dynamically manages network traffic through various RATs, achieving optimised transmission schemes to reduce latency, energy consumption, and communication costs. The framework is enhanced through xApp and rApp capabilities for real-time and non-real-time decision-making regarding optimal RAT selection and network configuration.

\subsection{Network Topology and Client Distribution}

The network topology considers client nodes distributed over a two-dimensional plane, mimicking realistic dynamic deployments. Each client connects to the FL server through the ORAN architecture, coordinating FL collaborations. Devices are dynamically assigned RATs based on network conditions, including congestion and link quality.

ORAN framework flexibility allows rApps and xApps to operate together, optimising network topology and resource allocation in both near-real-time and non-real-time scenarios. The rApp handles optimal RAT selection, prioritising power efficiency, while the xApp dynamically assigns channels and allocates physical resources based on rApp decisions, ensuring efficient bandwidth utilisation and minimal latency.

\subsection{Two-Stage optimisation Framework}

To optimise network resource usage for FL in ORAN networks, we present a coordinated two-stage approach leveraging rApp and xApp functionalities:

\subsubsection{Stage 1: RL-based RAT Selection (rApp)}
The rApp in Non-RT RIC employs Reinforcement Learning to optimise RAT selection and power allocation by monitoring network KPIs. The RL agent operates in state space $\mathbb{S}$ representing network conditions and action space $\mathbb{A}$ including RAT selection and transmission parameters.

\textbf{State Space}: $s_t = (U_t, L_t, Q_t, P_t) \in \mathbb{S}$ where:
- $U_t$: user distribution vector
- $L_t$: traffic load per RAT  
- $Q_t$: QoS metrics vector
- $P_t$: available power levels

\textbf{Action Space}: $a_t = (r_k, p_k) \in \mathbb{A}$ where $r_k \in \mathbb{R}$ is RAT assignment and $p_k \in \mathbb{P}$ is power level for user $k$.

The agent optimises policy $\pi(a|s)$ using reward function:
\begin{equation}
\label{eq: Rewards}
R_t = \alpha \cdot \eta_t + \beta \cdot T_t - \gamma \cdot L_t 
\end{equation}
where $\eta_t$ is energy efficiency, $T_t$ is throughput, $L_t$ is latency, and $\alpha, \beta, \gamma$ are balancing weights.

\subsubsection{Stage 2: CNN-based Resource Allocation (xApp)}
The xApp in Near-RT RIC facilitates near-real-time optimisation using CNN-based classification for resource allocation policies. Inspired by \cite{habib2024transformer,qazzaz2024machine}, this model-based approach predicts time-dependent network states and dynamically allocates resources using well-tuned classification techniques.

\begin{figure*}[t]
    \includegraphics[scale=0.29]{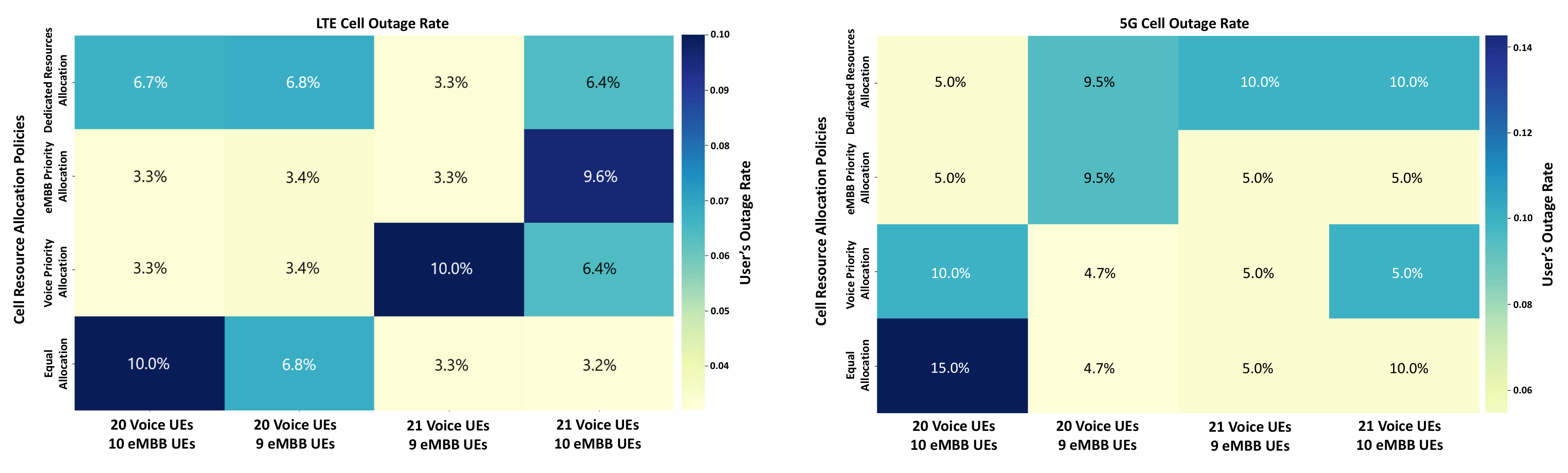}
    \centering
    \caption{Outage rates for multi-RAT network under different resource allocation policies.}
    \label{fig:outage probabilities for MPTCP and TCP protocols}
\end{figure*}

\textbf{CNN Architecture}: Input layer processes network features $(L_t, Q_t, r_t)$, hidden layers extract patterns, and output layer selects optimal policy $p^* \in \{P_1, P_2, P_3, P_4\}$.

\subsection{Dynamic Resource Allocation Policies}

We implement four resource allocation policies for distributing total resources $PRB_s$ across voice clients $(N_{voice})$ and eMBB clients $(N_{eMBB})$ including FL users:

\textbf{Equal Allocation}: Resources are divided equally without QoS consideration:
\begin{equation}
PRB_{client} = \frac{PRB_s}{N_{voice} + N_{eMBB}}  
\end{equation}

\textbf{Voice Priority}: Provides voice clients with $M$ times additional resource allocation, as follows:
\begin{equation}
PRB_{eMBB} = \frac{PRB_s}{(M \cdot N_{voice}) + N_{eMBB}}, 
\end{equation}
\begin{equation}
    \quad PRB_{voice} = M \cdot PRB_{eMBB}
\end{equation}
\textbf{eMBB Priority}: Gives eMBB clients priority by allocating $K$ times more resources than other users to handle high data traffic demands:
\begin{equation}
PRB_{voice} = \frac{PRB_s}{N_{voice} + (K \cdot N_{eMBB})}, 
\end{equation}
\begin{equation}
    \quad PRB_{eMBB} = K \cdot PRB_{voice}
\end{equation}

\textbf{Dedicated Reservation}: This policy reserves specific resource portions ($\alpha_v$ for voice users and $\beta_e$ for eMBB users) from the total available PRBs:
\begin{equation}
PRB_{voice} = \frac{\alpha_v \cdot PRB_s}{N_{voice}}, 
\end{equation}
\begin{equation}
    \quad PRB_{eMBB} = \frac{\beta_e \cdot PRB_s}{N_{eMBB}}
\end{equation}

\begin{algorithm}[t]
\caption{EcoFL optimisation Framework}
\label{algo:FedORA}
\begin{algorithmic}[1]
\Require Network state $s_0$, FL model $w_0$
\State \textbf{Initialize:} RL policy $\pi$, CNN model $\theta$
\For{each FL round $t = 1, 2, \ldots, T$}
    \State \textbf{// Stage 1: rApp RAT Selection}
    \State Observe network state $s_t$
    \For{each client $k \in K$}
        \State Select RAT: $r_k = \pi(s_t, k)$
        \State Assign power: $p_k = \pi_{power}(s_t, k)$
    \EndFor
    \State Execute actions, observe rewards $R_t$
    \State Update RL policy: $\pi \leftarrow \pi + \alpha \nabla_\pi \log \pi(a_t|s_t) R_t$
    
    \State \textbf{// Stage 2: xApp Resource Allocation}
    \State Extract features: $f_t = (L_t, Q_t, r_t)$
    \State Predict policy: $p^* = \text{CNN}(f_t; \theta)$
    \State Allocate resources according to $p^*$
    
    \State \textbf{// FL Training}
    \For{each selected client $k$}
        \State Download global model $w_t$
        \State Perform local training: $w_k^{t+1} = \text{LocalUpdate}(w_t, \mathbb{D}_k)$
        \State Upload model update
    \EndFor
    \State Aggregate updates: $w_{t+1} = \frac{1}{K} \sum_{k=1}^K w_k^{t+1}$
\EndFor
\end{algorithmic}
\end{algorithm}

These policies provide the xApp with flexible resource management strategies that can be dynamically selected based on network conditions and FL training requirements. The CNN-based xApp analyses current network state features, including traffic load, QoS metrics, and selected RATs, to determine the optimal policy for each time interval. This adaptive policy selection ensures that FL clients receive adequate resources during training phases while maintaining service quality for other network users. The effectiveness of each policy depends on the specific network scenario, with Equal Allocation providing simplicity, Priority policies addressing service-specific requirements, and Dedicated Reservation ensuring guaranteed resource availability for critical services.

% \subsection{Energy Consumption Model}

% Energy efficiency is achieved through RL-based power management with Power Plan IDs $(P_{ID})$ representing discrete power configurations: Low $(P_S)$, Mid $(P_M)$, and Full $(P_F)$ power modes.

% Total network power consumption:
% \begin{equation}
% \label{eq:power_total}
% P_{total} = \sum_{m=1}^{|K|} P_{ID,m}^{(r)}
% \end{equation}

% Energy efficiency metric:
% \begin{equation}
% \label{eq:efficiency}
% \eta_{ee} = 1 - \frac{P_{total}}{|K| \cdot P_F^{(r)}}
% \end{equation}

% Server-side energy cost:
% \begin{equation}
% \label{eq:server_energy}
% E_s(x_s, \mathbf{y}_s) = x_s \cdot E_s^{base} + \sum_{a \in \mathcal{A}} \sum_{r \in \mathcal{R}} y_{r,a,s} \cdot e_{a,s}
% \end{equation}
% where $x_s$ indicates server activation, $E_s^{base}$ is baseline energy, and $e_{a,s}$ represents incremental energy cost for application $a$ on server $s$.

\subsection{Energy Consumption Model}

Energy efficiency is a critical objective in EcoFL, achieved through comprehensive power management across both client devices and network infrastructure. Our energy model encompasses three primary components: client-side transmission power, network infrastructure consumption, and server-side computational costs.

\subsubsection{Client Power Consumption Model}
The client-side power consumption is governed by RL-based power management with Power Plan IDs $(P_{ID})$ representing discrete power configurations: Low $(P_S)$, Mid $(P_M)$, and Full $(P_F)$ power modes. For client $m$ using RAT $r$, the instantaneous power consumption is:

\begin{equation}
\label{eq:client_power}
P_{m}^{(r)}(t) = P_{ID,m}^{(r)} \cdot \alpha_{m}^{(r)} + P_{idle,m}^{(r)}
\end{equation}
where $\alpha_{m}^{(r)} \in [0,1]$ represents the transmission activity factor and $P_{idle,m}^{(r)}$ is the idle power consumption.

The total network power consumption across all clients is:
\begin{equation}
\label{eq:power_total}
P_{total}(t) = \sum_{m=1}^{|K|} P_{m}^{(r)}(t) = \sum_{m=1}^{|K|} \left( P_{ID,m}^{(r)} \cdot \alpha_{m}^{(r)} + P_{idle,m}^{(r)} \right)
\end{equation}

\subsubsection{Energy Efficiency Optimization}
The energy efficiency metric quantifies the relative power savings compared to worst-case full power operation:
\begin{equation}
\label{eq:efficiency}
\eta_{ee}(t) = 1 - \frac{P_{total}(t)}{|K| \cdot P_F^{(r)}}.
\end{equation}

To optimise energy consumption while maintaining QoS, we formulate the power allocation problem as:
\begin{equation}
\label{eq:power_optimization}
\begin{aligned}
\max_{\{P_{ID,m}\}} & \quad \eta_{ee}(t) \\
\text{s.t.} & \quad R_m^{(r)} \geq R_{min,m}, \quad \forall m \in K \\
& \quad L_m^{(r)} \leq L_{max,m}, \quad \forall m \in K \\
& \quad P_{ID,m}^{(r)} \in \{P_S, P_M, P_F\}, \quad \forall m
\end{aligned}
\end{equation}
where $R_m^{(r)}$ and $L_m^{(r)}$ represent achievable data rate and latency for client $m$ on RAT $r$, with corresponding minimum rate and maximum latency constraints.

\subsubsection{RAT-Specific Power Models}
Different RATs exhibit distinct power consumption characteristics. For 5G NR and LTE, the power consumption can be modeled as:

\begin{equation}
\label{eq:rat_power_5g}
P_{5G}^{(m)} = P_{RF}^{(5G)} + P_{BB}^{(5G)} \cdot \rho_m + P_{PA}^{(5G)} \cdot \left(\frac{P_{tx,m}}{P_{max}^{(5G)}}\right)^{\gamma}
\end{equation}

\begin{equation}
\label{eq:rat_power_lte}
P_{LTE}^{(m)} = P_{RF}^{(LTE)} + P_{BB}^{(LTE)} \cdot \rho_m + P_{PA}^{(LTE)} \cdot \left(\frac{P_{tx,m}}{P_{max}^{(LTE)}}\right)^{\beta}
\end{equation}
where $P_{RF}$, $P_{BB}$, and $P_{PA}$ represent RF frontend, baseband processing, and power amplifier consumption respectively, $\rho_m$ is the resource utilization factor, and $\gamma$, $\beta$ are RAT-specific efficiency exponents.

% \subsubsection{Infrastructure Energy Model}
% Server-side energy cost encompasses both activation overhead and application-specific consumption:
% \begin{equation}
% \label{eq:server_energy}
% E_s(x_s, \mathbf{y}_s) = x_s \cdot E_s^{base} + \sum_{a \in \mathcal{A}} \sum_{r \in \mathcal{R}} y_{r,a,s} \cdot e_{a,s}
% \end{equation}
% where $x_s \in \{0,1\}$ indicates server activation, $E_s^{base}$ is baseline energy, and $e_{a,s}$ represents incremental energy cost for application $a$ on server $s$.

% The total infrastructure energy consumption includes RAN elements:
% \begin{equation}
% \label{eq:infrastructure_total}
% E_{infr}(t) = \sum_{s \in \mathcal{S}} E_s(x_s, \mathbf{y}_s)+\sum_{b \in \mathcal{B}}\left( P_{BS,b}^{base} + \sum_{r\in\mathcal{R}} P_{BS,b}^{(r)} \cdot\xi_{b,r}(t) \right)
% \end{equation}
% where $P_{BS,b}^{base}$ is base station $b$'s baseline power, $P_{BS,b}^{(r)}$ is RAT-specific power, and $\xi_{b,r}(t)$ represents the activation status of RAT $r$ at base station $b$.

\subsubsection{Infrastructure Energy Model}
Let $\mathcal{S}$, $\mathcal{A}$, $\mathcal{B}$, and $\mathcal{R}$ denote the sets of servers, applications, base stations, and radio access technologies, respectively.

Server energy consumption includes both fixed activation costs and variable application-specific consumption:
\begin{equation}
\label{eq:server_energy}
E_s(x_s, Y_s) = x_s \cdot E_s^{base} + \sum_{a \in \mathcal{A}} \sum_{r \in \mathcal{R}} y_{r,a,s} \cdot e_{a,s}
\end{equation}
where $x_s \in \{0,1\}$ indicates whether server $s$ is active, $E_s^{base}$ is the baseline energy consumption when active, $Y_s = \{y_{r,a,s}\}$ represents the allocation variables, and $e_{a,s}$ is the energy consumption rate for application $a$ on server $s$.

Total infrastructure energy consumption combines server and RAN components:
\begin{equation}
\label{eq:infrastructure_total}
E_{infr} = \sum_{s \in \mathcal{S}} E_s(x_s, Y_s) + \sum_{b \in \mathcal{B}}\left( P_{BS,b}^{base} + \sum_{r\in\mathcal{R}} P_{BS,b}^{(r)} \cdot\xi_{b,r} \right)
\end{equation}
where $P_{BS,b}^{base}$ is base station $b$'s baseline power consumption, $P_{BS,b}^{(r)}$ is the additional power for RAT $r$, and $\xi_{b,r} \in \{0,1\}$ indicates whether RAT $r$ is active at base station $b$.

\subsubsection{FL-Specific Energy Considerations}
During FL training rounds, additional energy costs arise from model computation and communication. The energy consumed for FL training at client $m$ during round $t$ is:
\begin{equation}
\label{eq:fl_energy}
E_{FL,m}(t) = E_{comp,m} \cdot \tau_{local} + E_{comm,m}^{up} + E_{comm,m}^{down}
\end{equation}
where $E_{comp,m}$ is computational energy per local epoch, $\tau_{local}$ is the number of local epochs, and $E_{comm,m}^{up}$, $E_{comm,m}^{down}$ represent uplink and downlink communication energy costs.

The communication energy depends on the selected RAT and model size:
\begin{equation}
\label{eq:comm_energy}
E_{comm,m}^{up} = \frac{|\theta| \cdot P_{tx,m}^{(r)}}{R_m^{(r)}}, \quad E_{comm,m}^{down} = \frac{|\theta| \cdot P_{rx,m}^{(r)}}{R_m^{(r)}}
\end{equation}
where $|\theta|$ represents the model parameter size in bits, and $P_{tx,m}^{(r)}$, $P_{rx,m}^{(r)}$ are transmission and reception power levels.

This comprehensive energy model enables the RL agent to make informed decisions about RAT selection and power allocation, optimising the trade-off between energy efficiency and FL performance while ensuring QoS requirements are satisfied across all network services.

\section{Simulation Results and Discussion} 
\label{sec: Simulation and results}

We implemented our framework using Python with ORAN 7.2x split simulation. The setup includes offline model training for baseline establishment, followed by online testing. The simulation environment consists of one macro-cell (LTE eNB) and one micro-cell (5G NR gNB) serving 50 clients total.

Two traffic types are simulated: FL model updates (classified as eMBB) and voice traffic. QoS requirements align with 3GPP standards \cite{navarro2020survey}: voice traffic requires 0.1 Mbps throughput with 100 ms delay, while eMBB traffic requires 10 Mbps throughput with 80 ms delay. Table \ref{tab:simulation_setup} summarises simulation parameters.

\begin{table}[t]
    \centering
    \scriptsize
    \caption{Simulation Parameters}
    \label{tab:simulation_setup}
    \begin{tabular}{|p{2.45cm}|p{2.25cm}|p{2.25cm}|}
        \hline
        \textbf{Parameter} & \textbf{5G NR} & \textbf{LTE} \\
        \hline
        Bandwidth & 20 MHz & 10 MHz \\
        Carrier Frequency & 3.5 GHz & 800 MHz \\
        Subcarrier Spacing & 15 kHz & 15 kHz \\
        Max Tx Power & 43 dBm & 38 dBm \\
        Channel Model & 3GPP urban micro & 3GPP urban macro \\
        \hline
        \textbf{RL Parameters} & \multicolumn{2}{c|}{\textbf{rApp Configuration}} \\
        \hline
        Batch Size & \multicolumn{2}{c|}{32} \\
        Initial Steps & \multicolumn{2}{c|}{2000} \\
        Learning Rate & \multicolumn{2}{c|}{0.4} \\
        Discount Factor & \multicolumn{2}{c|}{0.92} \\
        \hline
        \textbf{CNN Parameters} & \multicolumn{2}{c|}{\textbf{xApp Configuration}} \\
        \hline
        Learning Rate & \multicolumn{2}{c|}{$10^{-2}$} \\
        Batch Size & \multicolumn{2}{c|}{32} \\
        Architecture & \multicolumn{2}{c|}{Input=4, Hidden=2$\times$6, Output=4} \\
        \hline
    \end{tabular}
\end{table}

Our optimisation strategy employs ORAN's RIC with rApp managing client communication via Deep Q-Learning for dynamic RAT adjustment and trained CNN (xApp) classifying optimal policies for resource management. The outer optimisation (rApp RL) iteratively updates Q-values to optimise network topology, while the inner optimisation (xApp CNN) manages network resources for effective model update sharing.

\begin{figure}[t]
    \centering
    \includegraphics[scale=0.33]{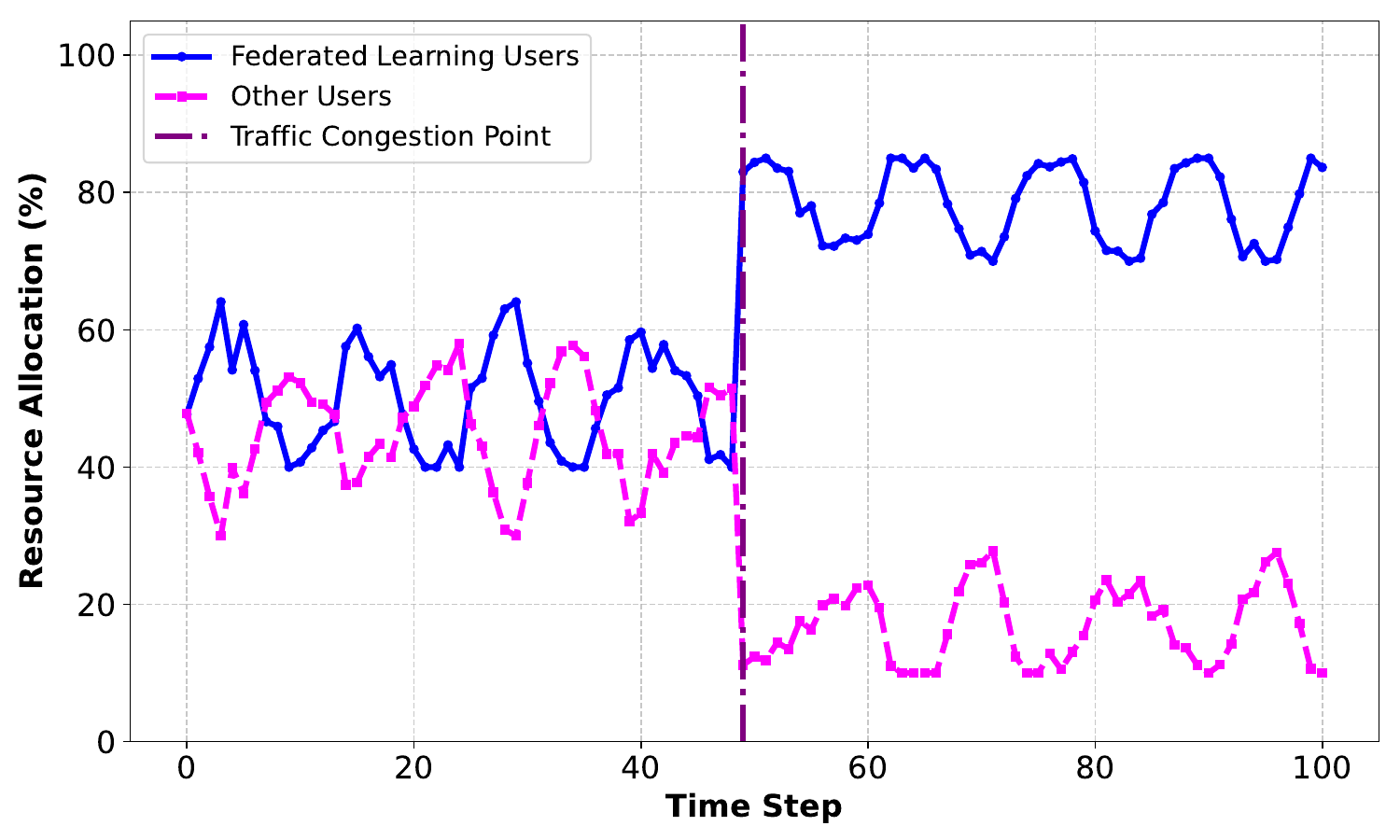}
    \caption{Dynamic resource allocation showing FL prioritisation during network congestion.}
    \label{fig:Resource allocation dynamics}
\end{figure}

The monitoring xApp in Near-RT RIC actively responds to escalating resource needs, enabling EcoFL to identify heightened RAN demands and intelligently shift allocation to favour FL users during training phases, as shown in Figure \ref{fig:Resource allocation dynamics}. This dynamic policy adoption significantly improves network communication within the ORAN architecture, ensuring intelligent optimisation of client participation without compromising critical services such as voice service.

\begin{figure}[!t]
    \centering
    \includegraphics[scale=0.42]{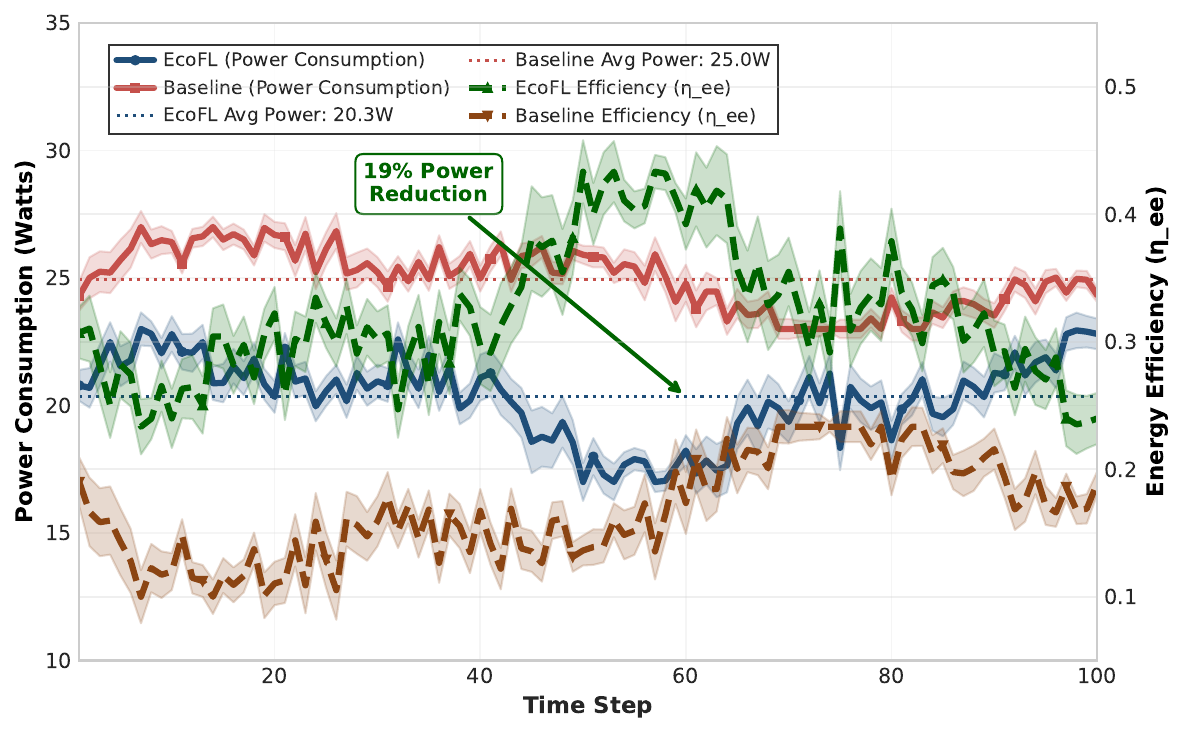}
    \caption{Energy performance and stability compared to baseline single-RAT approaches.}
    \label{fig:energy_performance}
\end{figure}

Figure \ref{fig:energy_performance} illustrates the energy performance comparison between EcoFL and baseline approaches over 100 time steps of FL training. The results demonstrate EcoFL's superior energy efficiency through dynamic multi-RAT selection and intelligent resource allocation.

The power consumption analysis reveals that EcoFL maintains an average power consumption of 20.08W compared to the baseline's 24.8W, achieving a consistent 19\% power reduction throughout the training process. This reduction is attributed to the RL-based RAT selection mechanism that intelligently chooses the most energy-efficient radio access technology based on real-time network conditions and FL training requirements.

The energy efficiency metric $\eta_{ee}$, calculated using Eq. (\ref{eq:efficiency}), demonstrates EcoFL's superior performance with an average efficiency of 0.330 compared to the baseline's 0.174, representing an 89.7\% improvement. The figure \ref{fig:energy_performance} illustrates that EcoFL not only consumes less power but also exhibits more stable performance with lower variance throughout the training period.

Our FL implementation leverages PyTorch APIs for CNN-based local training on the CIFAR-10 dataset (60,000 colour images across 10 categories). Each client trains locally on randomised data subsets, sharing only model updates to preserve privacy while iteratively improving global model accuracy.

Performance evaluation against leading FL algorithms (FedAvg \cite{mcmahan2017communicationefficient}, FLAIR \cite{sharma2023flair}, Greedy FL \cite{mehta2023greedy}) introduces realistic constraints with random client mobility in dynamic environments. Unlike existing models that often overlook communication challenges, our methodology directly assesses performance under potential intermittent connectivity.

\begin{table}[!t]
    \centering
    \scriptsize
    \setlength{\tabcolsep}{3.5pt}
    \renewcommand{\arraystretch}{1.1}
    \begin{tabular}{cccccccc}
        \toprule
        \textbf{Model} & \textbf{\shortstack{Local\\Algorithm}} & \textbf{Dataset} & \textbf{Accuracy} & \textbf{Loss} & \textbf{\shortstack{Std.\\Dev.}} & \textbf{\shortstack{FL\\QoS}} & \textbf{\shortstack{Avg. Power\\(Watts)}} \\
        \midrule
        EcoFL & CNN & CIFAR-10 & \textbf{96.5\%} & \textbf{0.038} & \textbf{0.15} & \checkmark & \textbf{20.08} \\
        \hline
        FLAIR & CNN & CIFAR-10 & 94.1\% & 0.042 & 0.32 & \sffamily $\times$ & 24.80 \\
        \hline
        FedAvg & CNN & CIFAR-10 & 85.3\% & 0.085 & 0.40 & \sffamily $\times$ & 24.80 \\
        \hline
        Greedy FL & CNN & CIFAR-10 & 90.2\% & 0.093 & 0.29 & \sffamily $\times$ & 24.80 \\
        \bottomrule
    \end{tabular}
    \caption{Performance comparison of EcoFL against baseline FL models.}
    \label{tab:DFL_MPTCP}
\end{table}

\begin{figure}[!t]
    \centering
    \includegraphics[width=1.01\linewidth, height=0.25\textheight]{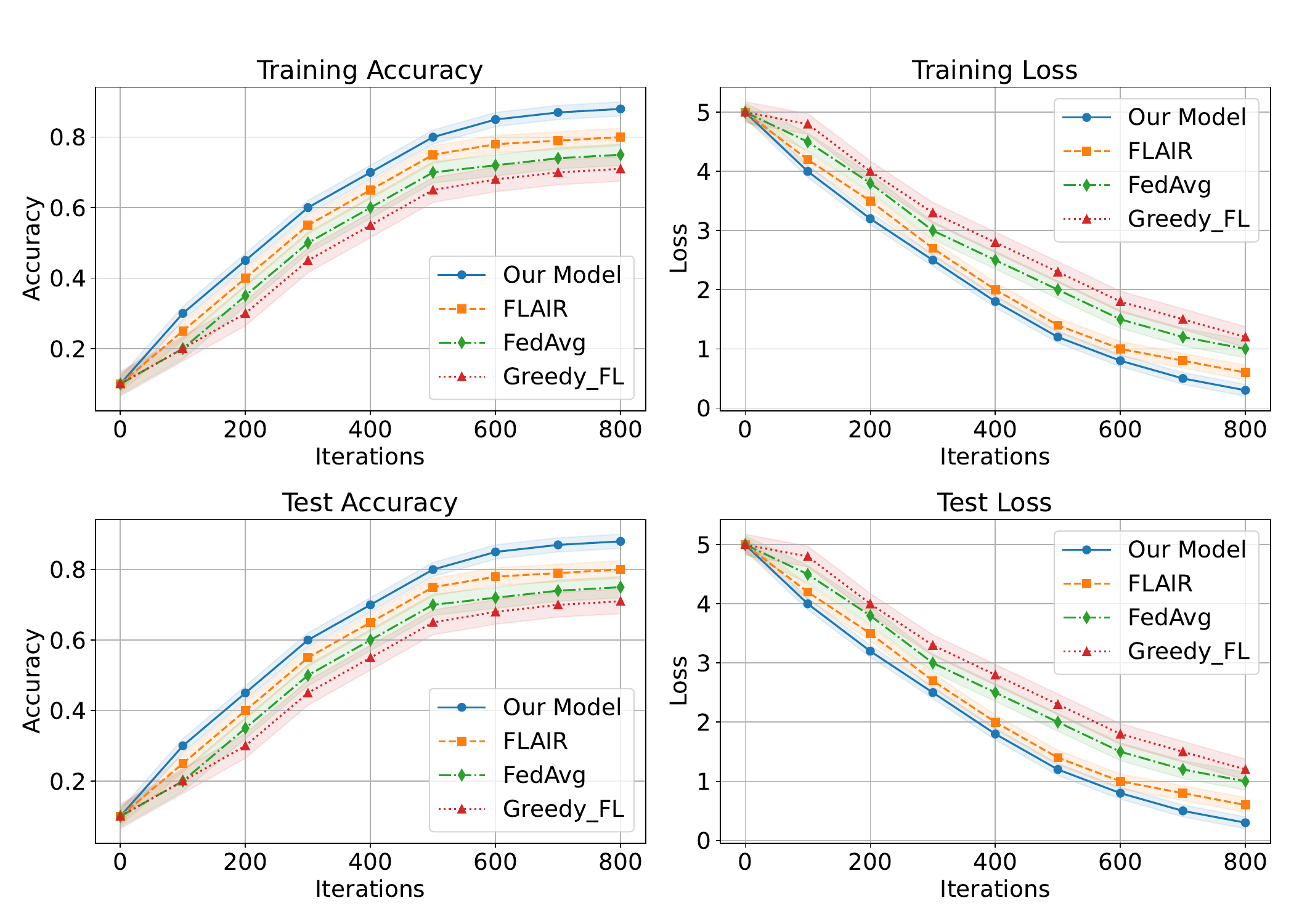}
    \caption{Comparative FL performance: accuracy and loss against baseline models.}
    \label{fig:MPTCP_accuracy}
\end{figure}

Results in Table \ref{tab:DFL_MPTCP} and Figure \ref{fig:MPTCP_accuracy} demonstrate that EcoFL achieves superior performance with 19\% lower average power consumption compared to fixed RAT scenarios. Multi-RAT and ORAN integration improve reliability through optimal policy adoption for low outage probability (Figure \ref{fig:outage probabilities for MPTCP and TCP protocols}). Our approach ensures effective energy and communication resource usage while maintaining higher client engagement and connectivity.

The enhanced resource allocation model improves real-world application requirements, including reliability and scalability, achieving robust model accuracy compared to traditional FL approaches. Energy consumption remains low through dynamic optimal policy selection that satisfies throughput and outage constraints for FL learning. This efficient resource management creates a comprehensive framework enabling effective resource utilisation, making it well-suited for scalable, collaborative learning applications requiring energy efficiency and network performance sensitivity.

\section{Conclusion} 
\label{sec: conclusion}

This work presents EcoFL, an innovative FL framework that leverages ORAN architecture to address privacy, communication efficiency, and energy consumption challenges in distributed learning. Our approach demonstrates substantial improvements in network performance, scalability, and system flexibility through multi-RAT support and intelligent resource management.

EcoFL achieves competitive accuracy (96.5\%) with 19\% lower power consumption compared to existing models, confirming its suitability for real-time applications requiring both efficiency and performance. The two-stage optimisation approach using RL-based RAT selection and CNN-based resource allocation effectively addresses communication bottlenecks in FL networks.

By leveraging rApp and xApp capabilities for dynamic optimisation in ORAN networks, our framework presents a robust solution that facilitates seamless collaborative learning while maintaining energy efficiency. This comprehensive approach enables effective resource utilisation, making it ideal for scalable applications that prioritise privacy, energy efficiency, and network performance.

% \section*{Acknowledgment}
% This research was funded by EPSRC CHEDDAR (EP/X040518/1), UKRI Grant EP/X039161/1, and MSCA Horizon EU Grant 101086218.

\bibliographystyle{IEEEtran}
\bibliography{references}

\end{document}